\documentclass[journal=cmatex,manuscript=article,layout=twocolumn]{achemso}

\usepackage[version=3]{mhchem} 
\usepackage{color}
\usepackage{gensymb}
\usepackage{tabularx}
\usepackage{array}
\usepackage{amsmath,amsfonts,amssymb,bm}
\usepackage{soul}

\author{Yanzhou Wang}
\affiliation{Beijing Advanced Innovation Center for Materials Genome Engineering, Department of Physics, University of
Science and Technology Beijing, Beijing 100083, China}
\alsoaffiliation{Department of Applied Physics, QTF Center of Excellence, Aalto University, FIN-00076 Aalto, Espoo, Finland}
\author{Zheyong Fan}
\affiliation{Department of Applied Physics, QTF Center of Excellence, Aalto University, FIN-00076 Aalto, Espoo, Finland}
\alsoaffiliation{College of Physical Science and Technology, Bohai University, Jinzhou, 121013, China}
\author{Ping Qian}
\affiliation{Beijing Advanced Innovation Center for Materials Genome Engineering, Department of Physics, University of
Science and Technology Beijing, Beijing 100083, China}
\email{qianping@ustb.edu.cn}
\author{Tapio Ala-Nissila}
\affiliation{Department of Applied Physics, QTF Center of Excellence, Aalto University, FIN-00076 Aalto, Espoo, Finland}
\alsoaffiliation{Interdisciplinary Centre for Mathematical Modelling and Department of Mathematical Sciences, Loughborough University, Loughborough, Leicestershire LE11 3TU, United Kingdom}
\email{tapio.ala-nissila@aalto.fi}
\author{Miguel A. Caro}
\affiliation{Department of Electrical Engineering and Automation, Aalto University, FIN-02150 Espoo, Finland}
\email{mcaroba@gmail.com}

\title
  {Structure and Pore Size Distribution in Nanoporous Carbon}

\begin{document}

\begin{tocentry}
\begin{center}
\includegraphics[width = 1.0\textwidth]{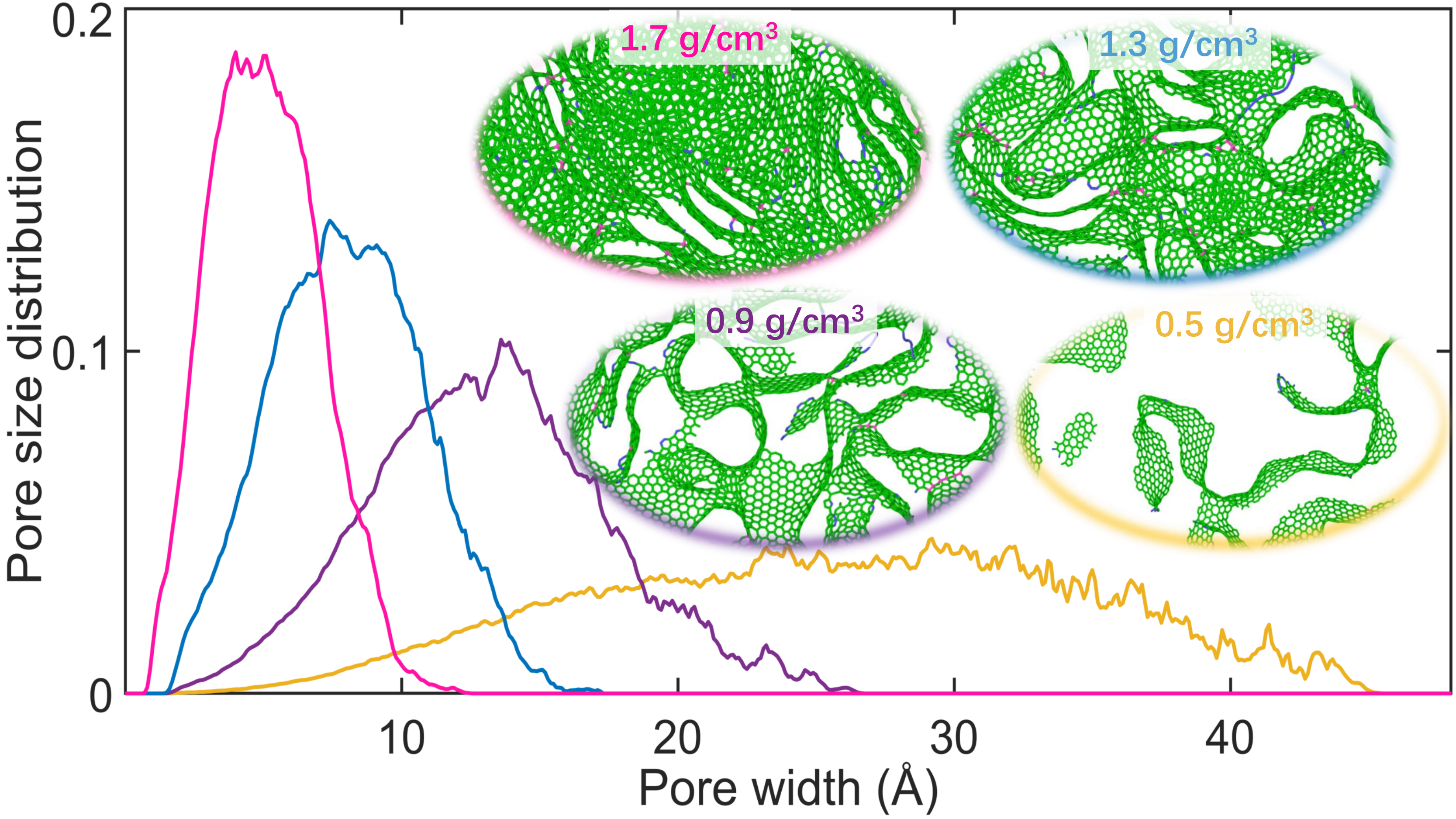}
\end{center}
\end{tocentry}

\begin{abstract}
We study the structural and mechanical properties of nanoporous (NP) carbon materials
by extensive atomistic machine-learning (ML) driven molecular dynamics (MD) simulations.
To this end, we retrain a ML Gaussian approximation potential (GAP) for carbon
by recalculating the a-C structural database of Deringer and Cs\'anyi
adding van der Waals interactions. Our GAP enables a notable
speedup and improves the accuracy of energy and force predictions.
We use the GAP to thoroughly study the atomistic structure and
pore-size distribution in computational NP carbon samples. These samples are generated
by a melt-graphitization-quench MD procedure over a wide range of densities (from 0.5 to
1.7~g/cm$^3$) with structures containing 131,072 atoms. Our results are in good
agreement with experimental data for the available observables, and provide a
comprehensive account of structural (radial and angular distribution functions, motif
and ring counts, X-ray diffraction patterns, pore characterization) and mechanical
(elastic moduli and their evolution with density) properties. Our results show relatively
narrow pore-size distributions, where the peak position and width of the distributions
are dictated by the mass density of the materials. Our data allow further work on computational
characterization of NP carbon materials, in particular for energy-storage applications,
as well as suggest future experimental characterization of NP carbon-based materials.
\end{abstract} 

\section{Introduction}

Nanoporous materials, where the pore-size distribution can be tuned according to growth
conditions, can be engineered for specific applications, such as ionic and
molecular transport~\cite{karger_2015}, biosensors~\cite{laurila_2017},
air and water purification~\cite{li_2020}, or energy storage~\cite{wang_2021}.
In particular, nanopores in disordered graphitic carbons arise
due to the misalignment and local curvature of the graphene-like sheets that make
up these materials. These pores are defined as any interstitial
space between graphene planes larger than the typical interlayer spacing in
graphite ($\approx$ 3.5~\AA{})~\cite{wang_2021}. When the pore sizes are of the order
of a few nanometers, the graphitic carbons are referred to as ``nanoporous'' (NP)
carbons. Nanopores significantly increase the effective surface area of NP carbons and confers
this class of carbon materials with the ability to intercalate a number of small
particles~\cite{vix_2005}. Most notably, the nanopores can accommodate ions which
can then migrate
through the material or intercalate/deintercalate upon application of an external
electric field, a mechanism exploited, for instance, in Li-ion batteries~\cite{wang_2021}.

In addition to the existence and utility of pores in NP carbons, graphitic carbons
in general display many attractive features that make them versatile materials for
a number of technological and industrial applications. They are chemically stable,
thanks to the large cohesive energy of the $sp^2$
bonds in elemental carbon. This makes them naturally resistant to corrosion
and other degradation processes affecting material lifetimes in devices where
electrochemical reactions are taking place, such as electrochemical sensors and
batteries~\cite{laurila_2017,wang_2021}. Graphitic carbons also display some astonishing
mechanical, thermal
and electronic properties, with graphene or carbon nanotubes (CNTs) topping the charts
of known materials in terms of some of their physical
properties~\cite{odom_1998,geim_2007,purohit_2014}.
At the same time,
carbon materials can be chemically functionalized to alter their reactivity,
an important example being graphene oxide (GO)~\cite{kumar_2013,savazzi_2018}.

While graphite itself
shows formidable in-plane strength, it can be easily deformed along the
direction of stacking of its constituent graphene monolayers. On the other hand,
disordered graphitic carbons, such as NP and glassy carbons (GCs), can show
isotropic mechanical response to dimensional changes in all spatial directions
even in the absence of noticeable $sp^3$ bonding. This happens whenever the graphitic
planes are randomly oriented such that
there is no preferential direction of stacking. These graphitic carbons are known
as ``non-graphitizing'', because it is not possible to experimentally transform them into crystalline
graphite even when applying large amounts of pressure and/or
heat~\cite{sundqvist_2021,wang_2021}.

For these reasons,
NP carbons have attracted a great deal of attention in recent years and in particular because
of their potential for electrochemical and electrocatalytical applications, and for
energy storage. Unfortunately, as is often the case for disordered materials,
experimental characterization of the structure of NP carbons is hindered by the
lack of mid- and long-range order, which renders common characterization techniques
inconclusive. The pore-size distribution in NP carbons is a critical parameter
to assess their potential for ion intercalation. To this end, in this work we
computationally examine the structure and pore-size distribution in NP carbons over a wide range of
mass densities using state-of-the-art machine-learning-driven molecular
dynamics (MLMD) simulations~\cite{deringer_2019}. Overcoming typical shortcomings
of accurate but expensive
density functional theory (DFT), on the one hand, and cheap but inaccurate
empirical interatomic potentials, on the other, MLMD grants us access to accurate
atomistic simulation on time and length scales previously out of
reach~\cite{caro_2018,muhli_2021}. In particular, we
carry out computational generation of NP structures, with approximately 130,000 atoms
each, in the mass density range from very low (0.5 g/cm$^3$) to relatively high (1.7
g/cm$^3$). The large number of atoms allows us to simulate
pores whose size is not limited by the dimensions of the simulation box. On these
structural models, we then perform a detailed characterization of the atomistic
structure, pore-size distribution and elastic properties. We expect that these
results will be useful in planning and
interpretation of experimental studies of NP carbons targeting specific pore
sizes, as well as to achieve a better fundamental understanding of the structure
of this important class of carbon materials.

\section{Methodology}

Nanoporous carbons can be experimentally synthesized by high-temperature heating
($\lessapprox$ 1800 K) of a carbon-rich precursor of either organic or inorganic
origin~\cite{wang_2021}. At these high temperatures,
the volatile molecules and elements in the precursor compound other than carbon
will evaporate away,
leaving behind the carbon-rich graphitic material. Since this is a very complicated
process to simulate directly (even with MLMD) due to the complex chemistry and time
scales involved,
in this study we pursue a different route. We prepare a highly disordered (liquid)
pure carbon precursor at very high temperature (9000~K) and then graphitize this system
at 3500~K using MLMD at different mass densities. This
approach~\cite{detomas_2016,detomas_2019}, which we review in
detail in this section, allows us to model the process of
graphitization within accessible simulation times and leads to the generation of
NP carbons with a wide range of pore sizes.

\subsection{GAP potential for carbon}

Our MLMD method of choice is the Gaussian approximation potential (GAP)
framework~\cite{bartok_2010,bartok_2015}. GAP is a kernel-based ML method for interpolation
of atomic potential energy surfaces, i.e., interatomic energies and forces.
To realistically simulate carbon graphitization we have retrained the GAP interatomic potential
for a-C of Deringer and Cs\'anyi~\cite{deringer_2017} (GAP17) for improved accuracy
and computational performance. We used the same structural database of
Ref.~\citenum{deringer_2017} as a baseline, but incorporated additional
graphite and dimer structures, and
recomputed it at the PBE+MBD level of theory~\cite{perdew_1996,tkatchenko_2012}
as implemented in VASP~\cite{bucko_2016,kresse_1996,kresse_1999}. The GAP17 database
contains dimers, amorphous and liquid carbon structures, amorphous surfaces, diamond, and
graphite configurations. The GAP potential itself uses kernel-based
regression~\cite{bartok_2010} with two-body (2b), three-body (3b)
and many-body (mb) descriptors.
Within this framework, a local atomic energy is predicted by the following expression:
\begin{align}
\bar{\epsilon}_* = &
\left( \delta^\text{(2b)} \right)^2 \sum_{s = 1}^{N_\text{sparse}^\text{(2b)}}
\sum_{j}'
\alpha_s^\text{(2b)} k^\text{(2b)}(\textbf{d}_{*j}^\text{(2b)}, \textbf{d}_s^\text{(2b)})
\nonumber \\
&+
\left( \delta^\text{(3b)} \right)^2 \sum_{s = 1}^{N_\text{sparse}^\text{(3b)}}
\sum_{jk}'
\alpha_s^\text{(3b)} k^\text{(3b)}(\textbf{d}_{*jk}^\text{(3b)}, \textbf{d}_s^\text{(3b)})
\nonumber \\
&+
\left( \delta^\text{(mb)} \right)^2 \sum_{s = 1}^{N_\text{sparse}^\text{(mb)}}
\alpha_s^\text{(mb)} k^\text{(mb)}(\textbf{d}_*^\text{(mb)}, \textbf{d}_s^\text{(mb)}),
\label{eq:gap}
\end{align}
where $k(\textbf{d}_*, \textbf{d}_s)$ is a kernel, measuring the similarity (with a score
from 0 to 1) between the atomic descriptor of the test configuration, $\textbf{d}_*$, and
the atomic descriptors in the database, $\{\textbf{d}_s\}_{s=1}^{N_\text{sparse}}$. Each
atomic environment, denoted with a star (*), has a set of 2b, 3b and mb descriptors
associated to it. Typically, this comprises a 2b descriptor for each atom pair it belongs to, a 3b
descriptor for each atomic triplet it belongs to, and one mb descriptor. The prime symbol
indicates that the sums over neighbors are carried out up to a certain cutoff.
The cutoff for the mb descriptor is embedded directly into it, since the mb descriptor
itself is constructed using truncated sums over neighbor pairs. Descriptors of
different types use different kernel functions, and have different sets of fitting
coefficients $\{\alpha_s\}_{s=1}^{N_\text{sparse}}$ (obtained during training)
and energy scale parameter
$\delta$ associated to them. Finally, $N_\text{sparse}$, which is
related\footnote{We differentiate between the \textit{training} set and the
\textit{sparse} set. The training set is all the data (atomic structures and target
properties) used to
train the model. The sparse set is a small subset of the descriptors found in the
training set, which is used to construct the kernel functions.} to
the number of training
configurations in the database, is also descriptor specific. Forces can be obtained
analytically from the gradients of Eq.~(\ref{eq:gap}). A detailed account of this
approach has been given by Bart\'ok and Cs\'anyi~\cite{bartok_2015} and in the
recent review by Deringer \textit{et al}~\cite{deringer_2021}. The retrained GAP
uses the same 2b and 3b descriptors as the original 
GAP17~\cite{deringer_2017} but replaces the SOAP mb descriptors~\cite{bartok_2013}
by \texttt{soap\_turbo} descriptors~\cite{caro_2019}. \texttt{soap\_turbo} offers
computational and accuracy advantages, such as faster computation of descriptors,
smoother radial basis functions and compression (a dimensionality reduction of the
descriptor resulting in significant speedup).

The longest range for interatomic
interactions that are accounted for by this GAP is 3.7~\AA{},
as in GAP17. Van der Waals
interactions are also explicitly incorporated, in a limited way,
via a tabulated long-range
pairwise interaction, which is optimized to reproduce the binding curve
of graphite, as previously done in Ref.~\citenum{rowe_2020}. We also used the
tabulated potential to model the strong exchange repulsion taking place at
very short interatomic distances; our ``core'' potential is fitted to reproduce
the dimer curve of carbon up to 1.8 keV (or, equivalently, down to 0.1~\AA{} separation).
The van der Waals and core-potential corrections were
absent from GAP17, and therefore we also expect this GAP to improve
these two aspects, relevant for graphitic
carbons and high-temperature or collision simulations, respectively.
Training of the new GAP was carried
out with the QUIP and GAP codes (with soap\_turbo libraries). All testing and
production MD simulations were done with the TurboGAP interface~\cite{ref_turbogap}.
The new GAP is freely available online~\cite{caro_2021}.

\begin{figure}[t]
    \centering
    \includegraphics[width = \columnwidth]{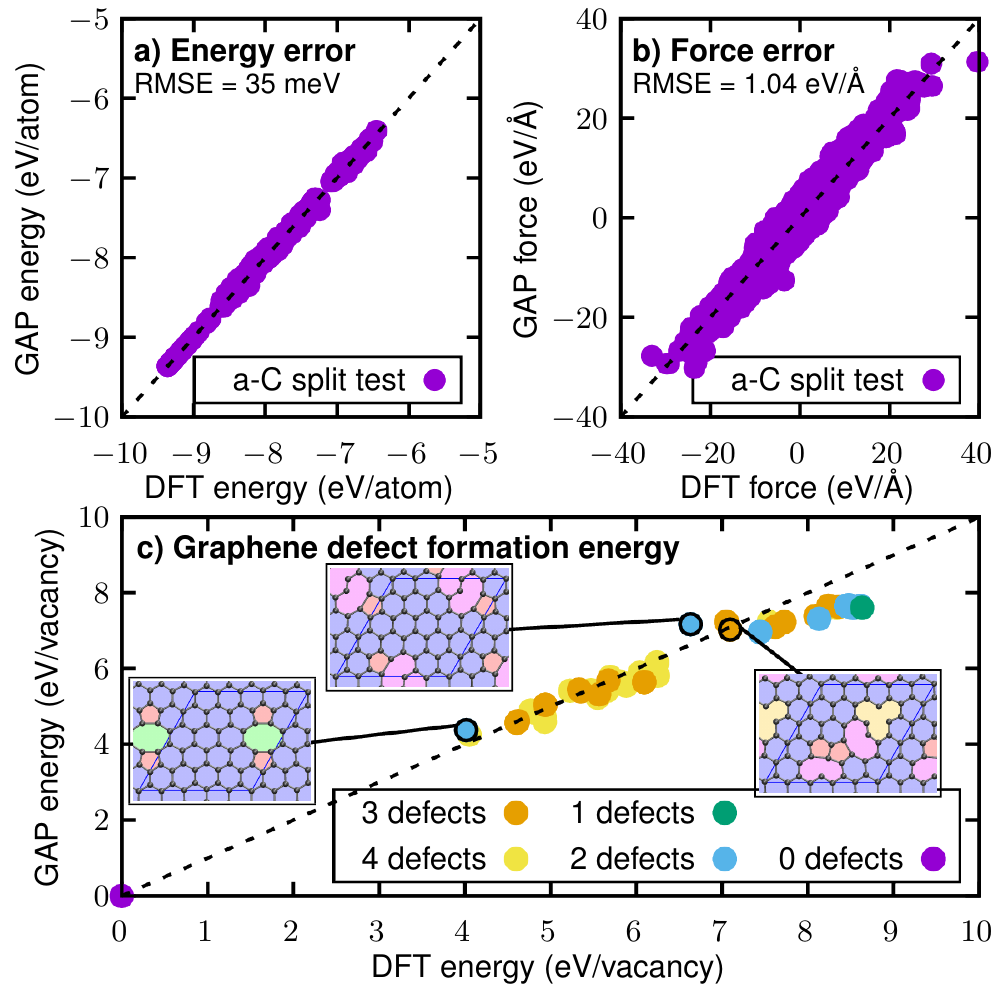}
    \caption{Accuracy benchmarks of the new carbon GAP. a) Scatter plot of
    predicted versus observed energies in a test set split from the training
    set. b) Same analysis as in a) but for the force components in this case.
    c) Comparison of 62 graphene structures with different numbers of vacancy
    defects computed with DFT versus the GAP prediction. The structures
    originally contained 50 atoms and vacancy defects were introduced by removing
    from 1 up to 4 atoms from random sites in the lattice. The structures are
    then relaxed and the DFT and GAP predictions compared. The insets show three
    example structures, where 5-, 6-, 8-, 9- and 12-membered rings are colored in
    red, blue, green, pink and yellow, respectively.}
    \label{fig:gap}
\end{figure}

To ensure that the new GAP can accurately simulate the graphitization
process, we performed some benchmarks as summarized in Fig.~\ref{fig:gap}.
We note that the root mean-square errors (RMSEs)
on the ``split test set'' configurations
(atomic configurations generated similarly to those used for training but that are
excluded from the fit and reserved for testing) are smaller than those of the
original GAP17. On the same test set, the new GAP reduces the error by 26\% and
9\% on energies and forces, respectively, as compared to GAP17.
A more application-specific accuracy test was run with the new GAP for the
vacancy defect formation energies in a graphene sheet. An accurate determination
of these energies is critical because the graphitization process consists
essentially of rearrangement of C atoms preferentially into 6-membered rings, with
any deviation constituting a defect and thus incurring an energy penalty. The
resulting ring and defect distribution in our porous carbon will thus be highly
sensitive to the GAP's ability to accurately reproduce these energies. In
Fig.~\ref{fig:gap}~c) we show how this GAP indeed succeeds at predicting the
relative vacancy formation energies of a series of representative defects in
close agreement with DFT. This is a stringent
transferability test since structures similar to these are not
included in the training set. The RMSE for defect formation energy
in graphene shown in Fig.~\ref{fig:gap}~c) are 0.52~eV/defect for
the new GAP and 1.41~eV/defect for GAP17. Based on our tests with
this and other GAPs, we ascribe this improved transferability to
the use of the new soap\_turbo descriptor, which provides smoother
kernels.

\begin{figure}[t]
    \centering
    \includegraphics[width = \columnwidth]{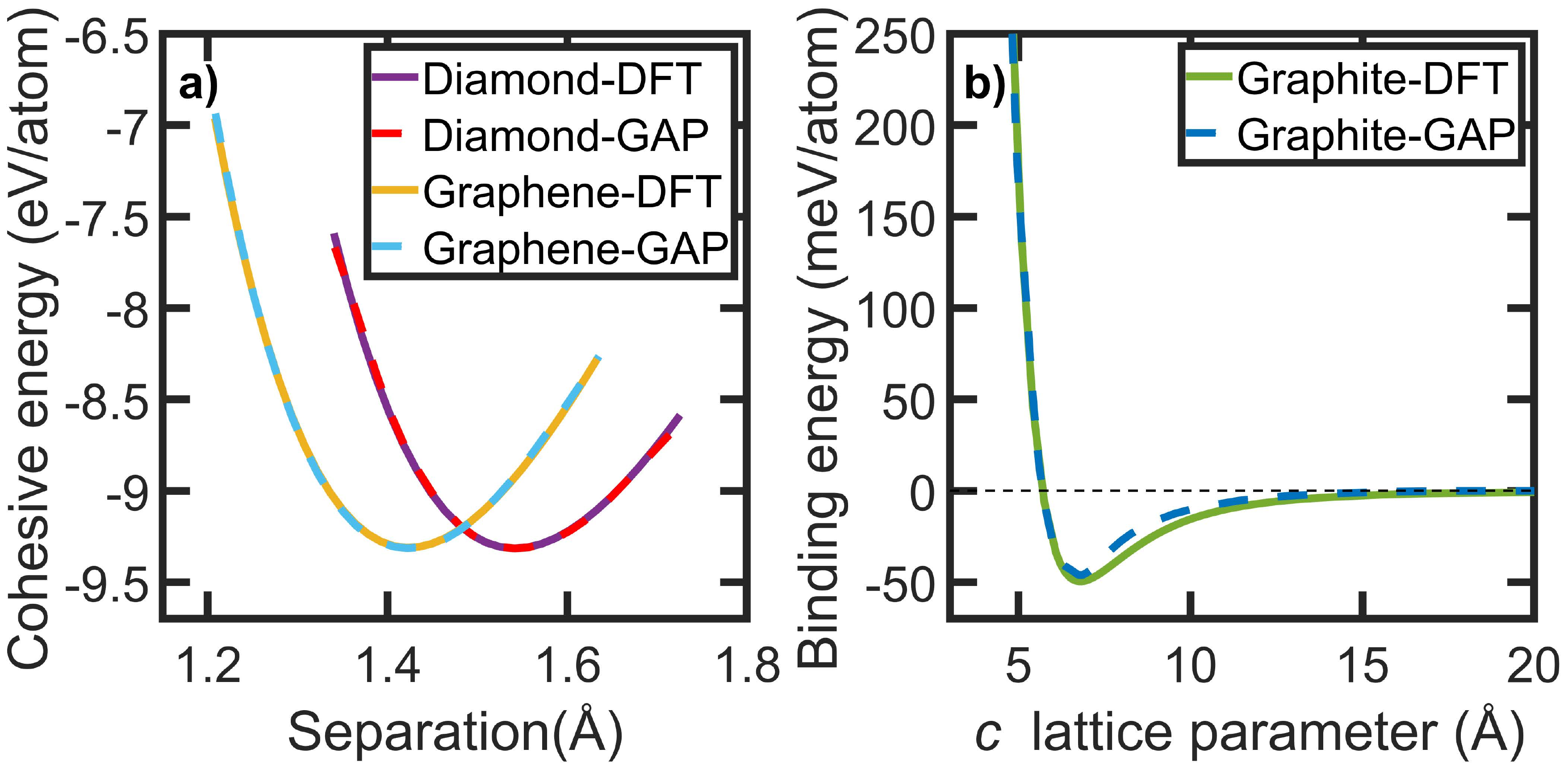}
    \caption{(a) Cohesive energy per atom versus the interatomic separation
    (first-nearest neighbors distance) for diamond and graphene.
    (b) Binding energy of graphene monolayers in graphite versus its lattice
    parameter $c$ (i.e., the exfoliation curve). The horizontal line is the
    reference for zero interaction.}
    \label{fig:E_V}
\end{figure}

The GAP also reproduces the cohesive energies
(Fig.~\ref{fig:E_V}), structural and elastic properties
(Table~\ref{table:bulk_proper}) of the three most important carbon allotropes:
diamond, graphite and graphene. We highlight the accurate description
of interlayer vdW interaction in graphite, with the only significant deviation
being the $C_{33}$ elastic constant. This is a particularly positive feature
of this potential given the notorious difficulty encountered by carbon force
fields to reproduce vdW binding~\cite{qian_2021}; PBE-DFT itself fails to bind
graphite at all in the absence of explicit vdW corrections.

\begin{table}[tb]
\centering
\caption{Lattice parameters (in \AA) and elastic constants (in GPa) of
diamond and graphite, calculated by DFT (PBE+MBD) and with the new GAP.}
\label{table:bulk_proper}
\begin{tabular}{lcccc}
\hline
\hline
 & \multicolumn{2}{c}{Diamond} & \multicolumn{2}{c}{Graphite} \\
 & DFT & GAP & DFT & GAP \\
\hline
 $a$ & $3.557$ & $3.560$ & $2.461$ & $2.463$ \\
 $c$ & & & $6.762$ & $6.803$ \\
 $C_{11}$ & $1071$ & $1028$ & $1083$ & $951$ \\
 $C_{12}$ & $134$ & $124$ & $163$ & $136$ \\
 $C_{13}$ & & & $-4$ & $4$ \\
 $C_{33}$ & & & $35$ & $129$ \\
 $C_{44}$ & $582$ & $538$ & $5$ & $7$ \\
 $C_{66}$ & & & $460$ & $407$ \\
\hline
\hline
\end{tabular}
\end{table}

A crucially important aspect of the MLMD approach concerns its computational efficiency. Here we
performed a comparison between the GAP17 running with
LAMMPS~\cite{plimpton_1995,ref_lammps} and the new GAP running
with TurboGAP for a test configuration containing 1000 C atoms (simple cubic
lattice with 2~g/cm$^3$ density and random initial velocities). We ran 1000 MD steps
on a 40 CPU-core machine (the same hardware for both codes). The TurboGAP
calculation ran in 0.47~CPUh or 1.7~ms/atom/MD step (42~s or 42~$\mu$s/atom/MD step
of walltime) whereas the LAMMPS calculation ran in 3.3~CPUh or 12~ms/atom/MD step
(297~s or 297~$\mu$s/atom/MD step of walltime), i.e., the new GAP
runs about seven times faster \textit{and} is more accurate.
There are three main reasons for this significant speedup. First,
the soap\_turbo descriptors are intrinsically faster than the original SOAP
because of more efficient iterative algorithms~\cite{caro_2019}. Second, soap\_turbo
implements compression, which in itself can be the single largest contribution
to computational speedup in kernel-based ML potentials~\cite{musil_2021}.
Third, TurboGAP implements efficient polynomial 3b kernels, which are also
significantly faster to compute than the squared-exponential 3b kernels used
in GAP17.

The result of adding the vdW
correction is a computational overhead that strongly depends on the cutoff
for the vdW interactions. For this benchmark comparison, a 10~\AA{} vdW cutoff,
the same as for the graphitization simulations in the next section, led to
$\approx 13$~\% slower TurboGAP execution (48~s, 0.53~CPUh or 1.9~ms/atom/MD
step). Therefore, with the new GAP and using the TurboGAP interface, it is possible to
tackle much larger systems than before which is of critical importance to achieve
realistic pore sizes in the present work.

\subsection{MD simulation details}

We carried out all the MD  simulations with
the new GAP interatomic potential for carbon and the TurboGAP
code~\cite{ref_turbogap}. For a comprehensive characterization of NP carbon,
we performed simulations
over a wide range of mass densities: $0.5$, $0.9$, $1.3$ and $1.7$~g/cm$^3$.
For each density value, a cubic simulation box with $131,072$ carbon atoms is
generated with randomized initial atomic positions. The side lengths of the
cubic boxes are $17.36$, $14.27$, $13.62$, and $11.54$~nm for the four densities
considered above, respectively. The simulation box dimensions give an idea of the
maximum pore sizes that can be modeled.

\begin{figure}[t]
    \centering
    \includegraphics[width = \columnwidth]{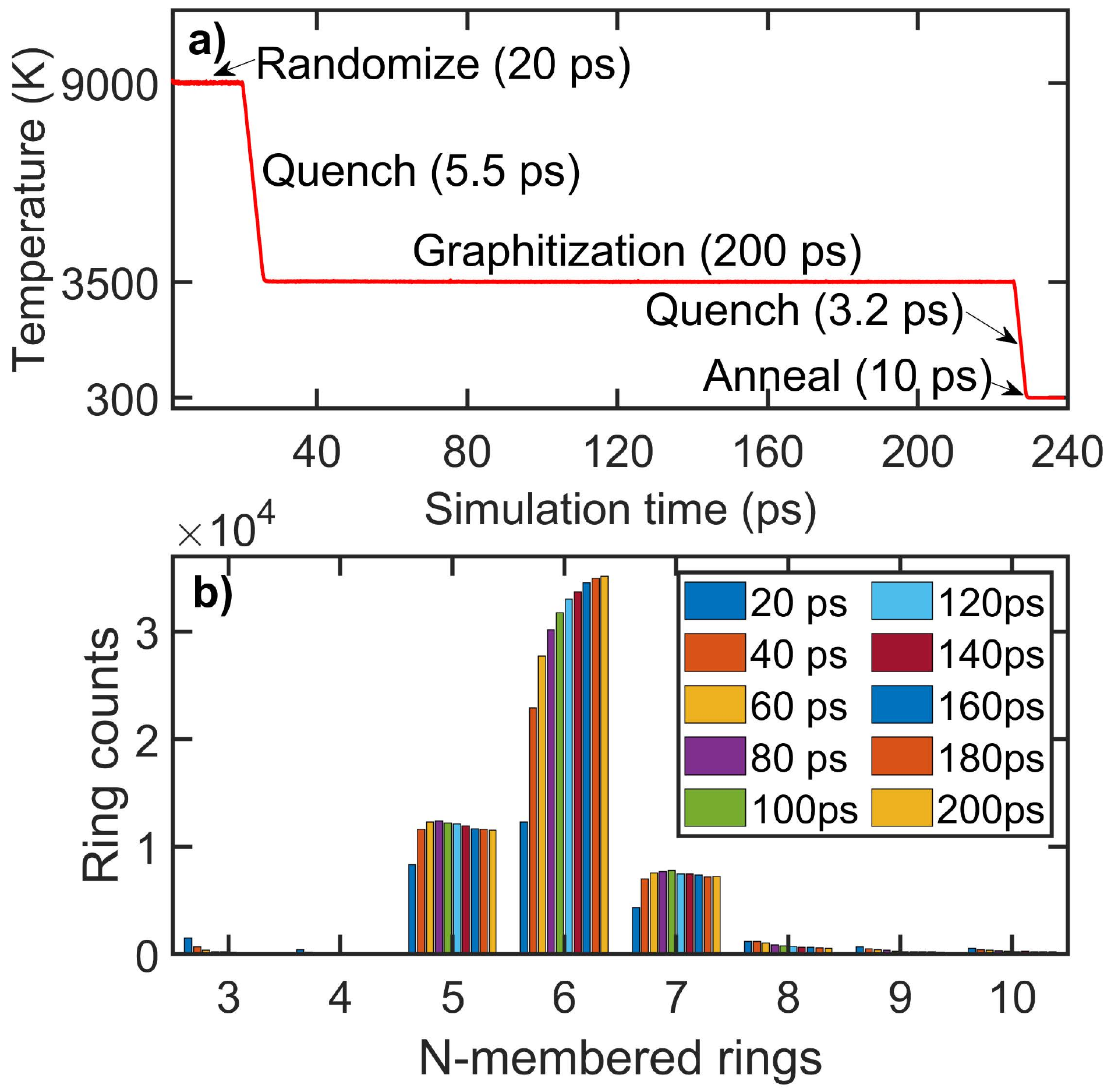}
    \caption{(a) Temperature profile during the melt-quench process. (b)
    Convergence of the number of $N$-membered rings ($3\leq N \leq 10$) during
    graphitization at $3500$~K in the system with density $0.5$ g/cm$^3$. The
    number of 6-membered rings initially grows very fast, significantly slowing
    down towards the 200~ps mark.}
    \label{fig:temp_prof_plus_ring_evol}
\end{figure}

The protocol for NP carbon generation
is as follows [see Fig.~\ref{fig:temp_prof_plus_ring_evol}(a)]. First, we hold
the highly disordered structure at a constant high temperature (9000 K) in the
liquid state for 20~ps. Second, we quench the liquid with a linear decaying temperature
profile for 5.5~ps down to 3500~K. The structure is then kept at 3500~K
for 200~ps. This is the graphitization step~\cite{detomas_2016,detomas_2019},
and the time has been specifically optimized to ensure that no significant further
graphitization takes place after this period by monitoring the creation of
6-membered rings as a function of time, which is negligible after 200~ps at
3500~K [see Fig.~\ref{fig:temp_prof_plus_ring_evol}(b)]. Finally, the graphitized
sample is quenched down to 300~K over 3.2~ps, and annealed at 300~K for 10~ps
to obtain the room-temperature structures.
The velocity-rescaling Berendsen thermostat~\cite{berendsen_1984} with time constant
$\tau_\text{t} = 100$~fs was used to control the temperature, and all GAP-MD
simulations were performed under periodic boundary conditions within fixed
box dimensions, with a Verlet integration time step of 1~fs. 

For each density we performed six independent MD simulations as described
above to assess the effect of individual random-seed structures on the
final results. This is not needed for studying short-range structural properties such
as coordination statistics, but is needed for accurate calculations of medium-
and long-range structural properties such as ring statistics and pore-size
distribution, as well as elasticity.

\subsection{Structural characterization}

After obtaining the MD trajectories we used OVITO~\cite{ovito_stukowski_2010} to
calculate the radial distribution function (RDF) and the angular distribution function
(ADF). The cutoff distance here was chosen as $2.0$, $1.9$, and $1.8$ \r{A} at
$9000$, $3500$, and $300$ K, respectively. This software was also used to render
the structural images and calculate the coordination numbers of the $sp$, $sp^2$
and $sp^3$ structural motifs using a cutoff distance of $1.8$ \r{A}, which fully
encloses the first-neighbors shell for NP carbon. We used a single frame to
calculate the RDF and ADF which is sufficient~\cite{Thomas_carbon_2016,Thomas_jncs_2019}
for systems with a large number of atoms.

The reciprocal-space diffraction scattering patterns, \textit{I}(\textit{Q}),
were calculated using the Debye scattering equation as implemented in the Debyer
code~\cite{ref_debye}:
\begin{equation}
 I(Q) = \sum_i \sum_j f_if_j \frac{\sin(Qr_{ij})}{Qr_{ij}}.
\label{debye}
\end{equation}
Here, $Q=4\pi\sin(\theta)/\lambda$ is the scattering parameter, where $\theta$ is the
diffraction half-angle, $\lambda$ is the wavelength (1.5406~\AA{} for the
commonly used Cu $K$-$\alpha$), $r_{ij}$ is the distance between
atoms $i$ and $j$, and $f$ is the atomic scattering factor for a single carbon atom.
As noted in Ref.~\citenum{Thomas_carbon_2016}, the minimum image convention causes
problems when using the Debye equation, and therefore it is necessary to map the
atoms into the primitive cell and treat the structure as an isolated cluster. 

The coordination numbers only account for short-range structural properties. To
study the medium-range structural properties, we calculated ring statistics,
using the shortest-path algorithm of Franzblau~\cite{ring_fran_prb_1991} as
implemented in the I.S.A.A.C.S. code~\cite{isaacs_le_2010}. We used the same cutoff
distances for nearest neighbors as in the calculation of the ADF. 

To quantitatively characterize the nanopore structures arising in our
samples, we calculated the size and shape distributions of the pores using
the zeo++ code~\cite{zeo_jmgm_2013}. For the pore size, first the three-dimensional
Voronoi network\cite{voro_netw_2012} is computed to establish the void space accessibility
within the pores, using the Voronoi decomposition method implemented in the Voro++
library~\cite{rycroft_2009}. We further take Monte Carlo samples to determine the
accessible volume, and eventually determine the radius of the largest sphere that encloses
the sampled point without overlapping any of the C atoms in the NP
structures. As a complement to pore size, we also present corresponding
pore-shape information using a stochastic ray-tracing algorithm~\cite{ray_trac_2013}
that shoots random rays through the accessible volume of the Voronoi cell
until they hit the wall of constituent C atoms. We take a probe radius
of 0.7 \AA{} (half of the typical bond length in the NP carbon samples)
and 130,000 points for Monte Carlo sampling.

\section{Results and discussion}

\subsection{Short-range order in liquid and NP carbon}

\begin{figure}
    \centering
    \includegraphics[width = \columnwidth]{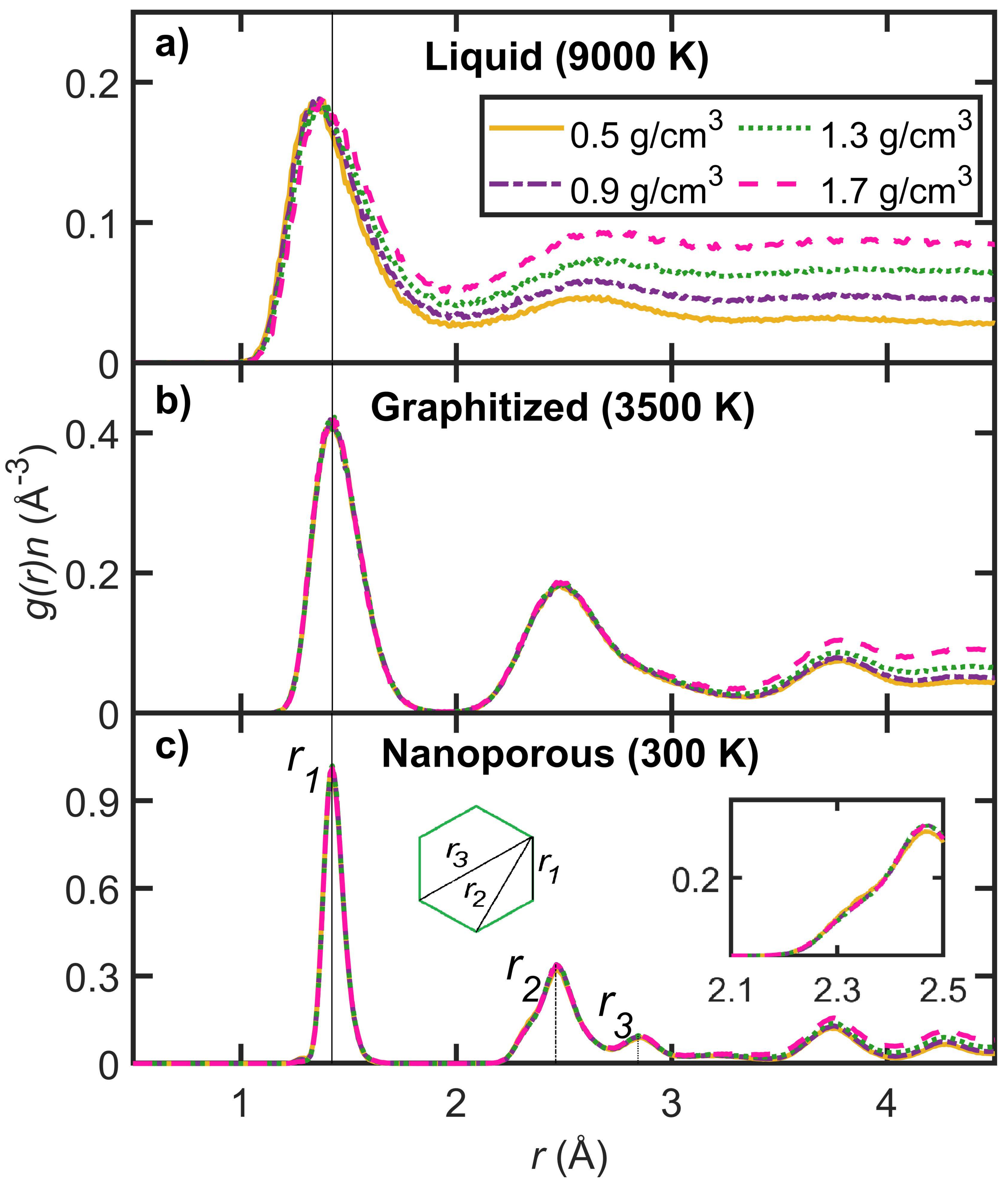}
    \caption{Density-scaled radial distribution functions $g(r)n$, where $n$ is
    the particle density, for (a) liquid, (b) graphitized, and (c) NP carbon
    structures. The insets in (c) give i) the illustration for the first three peak
    positions in a graphene/graphite hexagon, and ii) a magnified view in the $r$ range from
    2.1 to 2.5 \AA{}, where a small shoulder at $r$ $\approx$~$2.32$~\AA{} indicates the presence
    of pentagons in the NP structure. The vertical lines mark the first three
    nearest-neighbor distances of $1.42$ ($r_1$), $2.46$ ($r_2$) and $2.84$~\AA{}
    ($r_3$).}
    \label{fig:RDF}
\end{figure}

\begin{figure}
    \centering
    \includegraphics[width = \columnwidth]{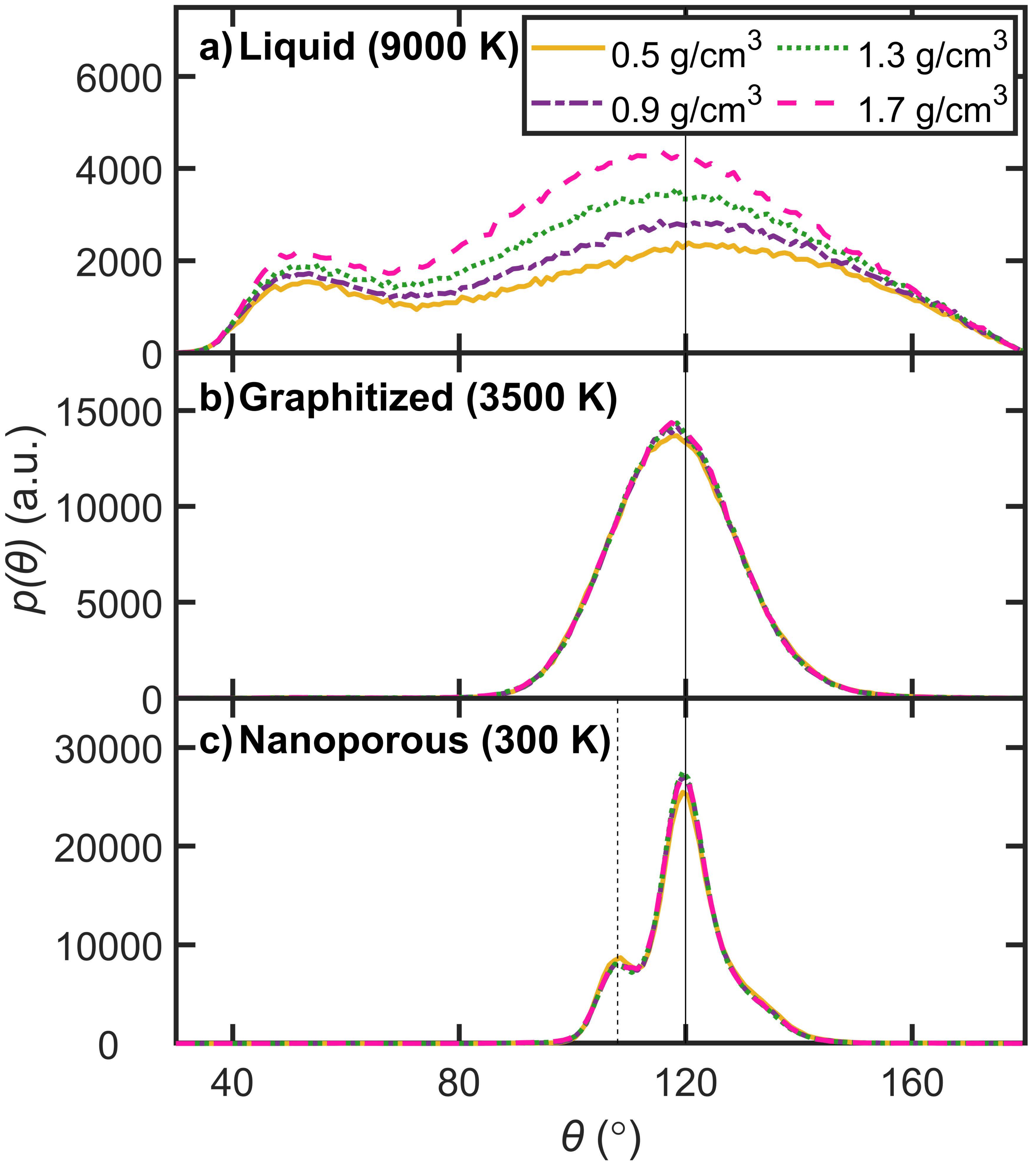}
    \caption{Angular distribution functions for (a) liquid, (b) graphitized, and
    (c) NP carbon structures. The solid and dashed vertical lines represent the
    angles for 120$\degree$ hexagons and 108$\degree$ pentagons,
    respectively.}
    \label{fig:ADF}
\end{figure}

To characterize the short-range order, we present the RDFs for the liquid (9000~K),
graphitized (3500~K) and NP (300~K) structures generated during the melt-quench
process in Fig.~\ref{fig:RDF}. We consider the product of RDF
$g(r)$ and density of particles \emph{n} to highlight the comparison between the
different density systems. The RDFs converge to their respective average density
of particles $n$ ($0.025$, $0.045$, $0.065$ and $0.085$~\AA$^{-3}$ for $0.5$,
$0.9$, $1.3$ and $1.7$ g/cm$^3$, respectively) as the interatomic distance $r$ tends
to infinity. 

Similar to the results obtained by GAP17 \cite{deringer_2017}, the liquid
structures in Fig.~\ref{fig:RDF}(a) are less strongly ordered and more diffuse,
showing a nonzero minimum at $\approx$~$2.0$ \AA, whereas the solid phases
in Fig.~\ref{fig:RDF}(b) and (c) are more ordered even in the medium-range
regions, exhibiting more individual peaks. These solid phases, especially at
300~K, display a clear gap between their first ($r_1 \approx$~$1.42$ \AA) and
second ($r_2 \approx$~$2.46$ \AA) peaks. Compared to the solid structures, in the high-temperature
liquid in Fig.~\ref{fig:RDF}(a) the peak shifts to a smaller $r$ ($\approx$~$1.35$ \AA),
which is also associated with the presence of
a bimodal angular distribution
in the corresponding ADFs of Fig.~\ref{fig:ADF}(a). 

After the 3500~K structure is quenched down to room temperature, the peak at
$r_3 \approx$~$2.84$~\AA{} becomes clearly noticeable in Fig.~\ref{fig:RDF}(c),
which corresponds to the third nearest-neighbor distance
in a graphite hexagon. That the first three peak positions show a close
correspondence to the first three neighbor distances
of graphite implies 6-membered rings are the dominant short-range structural
motif in these NP structures. Also, in the locally magnified view of
Fig.~\ref{fig:RDF}(c), a small shoulder at $\approx$~$2.32$~\AA{} (corresponding to
the second nearest-neighbor distance in a carbon pentagon) appears in the
early second nearest-neighbor region, a signature of the presence of some
5-membered rings.

To accompany the discussion with a coordination analysis of the first
nearest-neighbor regions of RDF above, we give ADFs in Fig.~\ref{fig:ADF}.
At 9000~K, in addition to the main peak at $\approx$ $120\degree$, an additional smaller
one appears at $\approx$~$53\degree$ (close to the 60\degree{} in an equilateral
triangle, Fig. \ref{fig:ADF}(a)), which means there are some irregular triplets
in existence in the liquid state. These irregular triplets might be the main
cause for the shifted RDF peak in \ref{fig:RDF}(a). We also note sizable
differences in their ADFs across the four sampled densities in the liquid state.
This is in contrast to the graphitized [Fig.~\ref{fig:ADF}(b)] and room-temperature
NP [Fig.~\ref{fig:ADF}(c)] samples, for which the coordination statistics show very
little dependence on the sample density. This is a clear sign that the short-range
order in NP carbon is rather independent of mass density.

For the liquid samples, the ADF values all tend to zero at $180\degree$, indicative
that there is almost no presence of flat linear $sp$-bonded chains. The most
stable linear $sp$ chains in \textit{solid} carbon are expected to show bond angles in the vicinity
of 155\degree~\cite{caro_2018c}, where the ADF in [Fig.~\ref{fig:ADF}(a)] is
still sizable. Our present result is inconsistent with previous findings for
simulated liquid carbon~\cite{deringer_2017,gap20_jcp_2020}, in which small-size
systems containing 216 atoms were modeled under period boundary conditions. Our tests
indicate that this discrepancy is due to small inaccuracies at very high
temperature for the energetics of linear chains as predicted by our GAP, which is
biased towards the more stable bent chains ($\sim 155$\degree) observed at low
temperature.

During the graphitization step, in Fig.~\ref{fig:ADF}(b), unstable triplets
rearrange into more stable motifs centered at $\approx$~$120\degree$. We can
clearly see the signature of near-complete graphitization, which mirrors the main
structural feature of graphite hexagons. Notably, in quenched NP carbon
[Fig.~\ref{fig:ADF}(c)], there is a new minor twin peak at 108$\degree$, which
corresponds to pentagonal motifs. This is strong evidence, further supported by
the ring analysis carried out later in this section, of the presence of
defective 5-membered rings in quenched NP carbon. Pentagon and heptagon units
play an important role in curving graphene fragments in NP and glassy
carbons~\cite{martin_prl_2019}.

\subsection{Morphology and short- and medium-range characterization of NP structures}

\begin{figure*}[t]
    \centering
    \includegraphics[width = 1.0\textwidth]{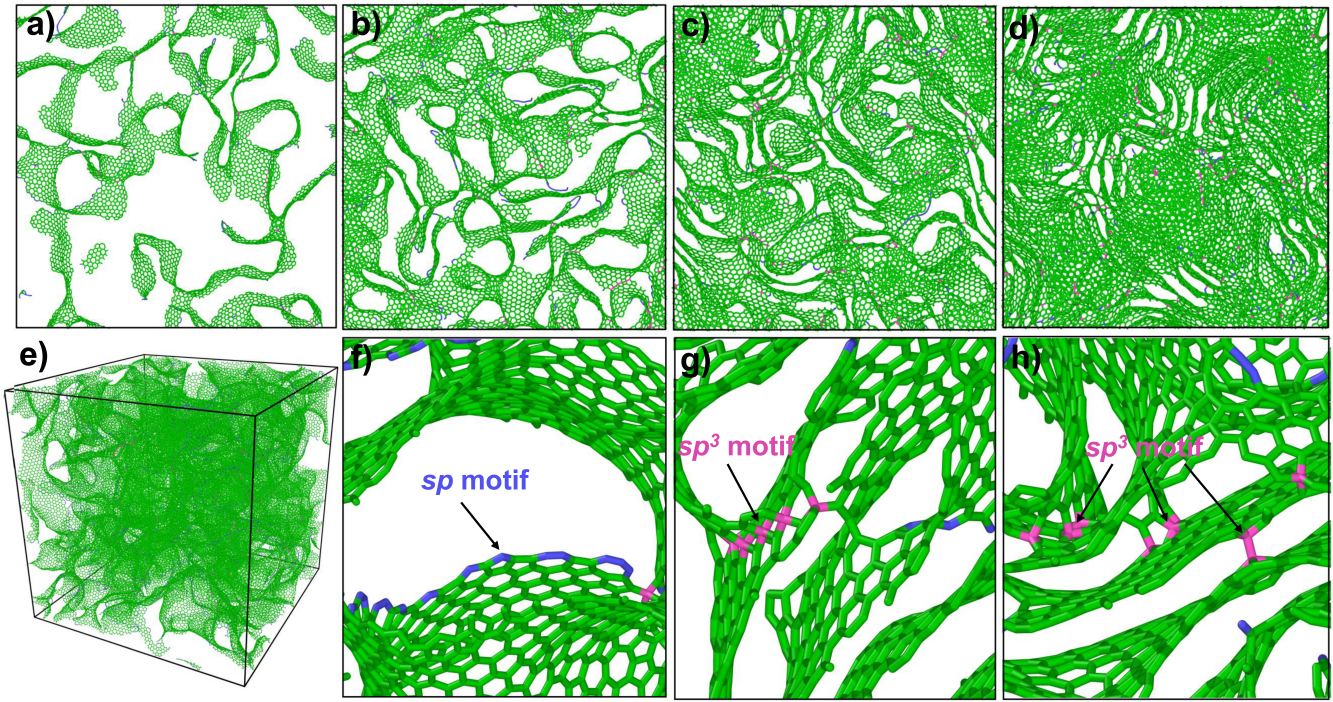}
    \caption{Cross-sectional slices ($1.5$ nm in thickness) of typical snapshots of
    NP carbon at the densities of (a) 0.5, (b) 0.9, (c) 1.3 and (d) 1.7 g/cm$^3$,
    as well as a (e) 3D snapshot of 0.5 g/cm$^3$. Indicated (f) $sp$ and
    (g, h) $sp^3$ motifs in the network of NP carbon. The $sp$, $sp^2$ and $sp^3$
    motifs are rendered in blue, green and red, respectively. Bonds with lengths
    shorter than $1.8$~\AA{} are displayed.}
    \label{fig:stru_slic}
\end{figure*}

\begin{table}[tb]
    \centering
    \begin{tabular}{p{1.5cm} p{1.5cm} p{1.5cm} p{1.5cm}}
    \hline
    \hline
    Density & $sp$ (\%)  & $sp^2$ (\%) & $sp^3$ (\%) \\
    \hline
    $0.5$ & $3.00$ ($0.10$) & $96.59$ ($0.10$) & $0.41$ ($0.03$) \\
    $0.9$ & $2.29$ ($0.08$) & $97.07$ ($0.06$) & $0.63$ ($0.06$) \\
    $1.3$ & $1.87$ ($0.06$) & $97.38$ ($0.05$) & $0.74$ ($0.03$) \\
    $1.7$ & $1.39$ ($0.04$) & $97.68$ ($0.09$) & $0.92$ ($0.05$) \\
     \hline
    \hline
    \end{tabular}
    \caption{Coordination fractions of $sp^n$ ($1\leq n \leq 3$) bonds
    versus different mass densities (g/cm$^3$). Standard deviations are
    given in parentheses.}
    \label{table:sp2}
\end{table}

\begin{figure}[t]
    \centering
    \includegraphics[width = \columnwidth]{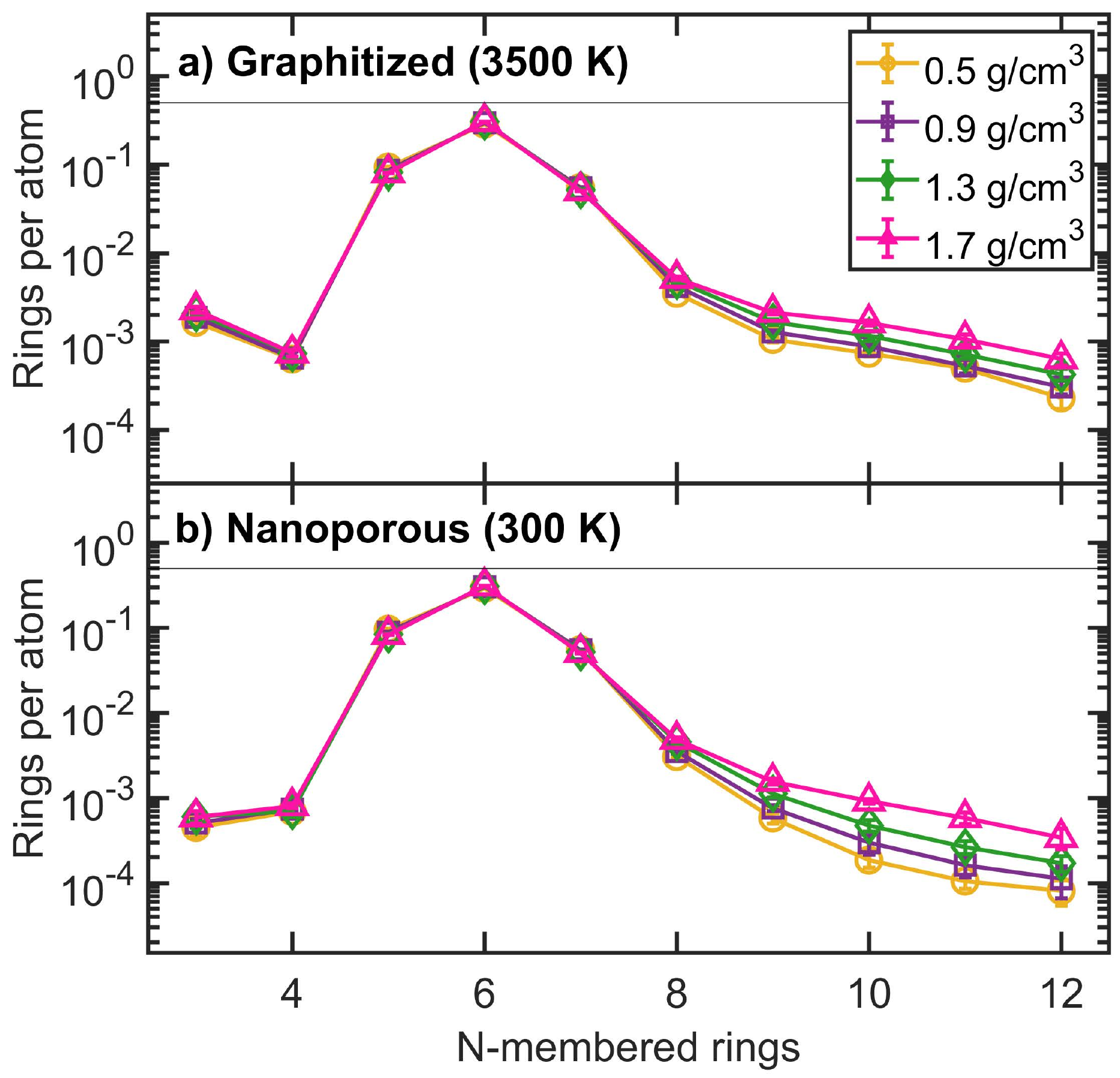}
    \caption{Number of rings per atom in (a) graphitized (3500 K) and (b)
    quenched (300 K) structures. The solid line indicates the reference
    value of 6-membered rings per atom in ideal bulk graphite. Standard
    deviations are given over six random seeds in simulations.}
    \label{fig:ring}
\end{figure}

Figure~\ref{fig:stru_slic} shows the cross-sectional morphology of our quenched NP
carbon samples for the four densities studied. Clearly, most of the carbon in these
samples makes up curved graphene fragments with predominantly $sp^2$ bonding, quantified
in Table~\ref{table:sp2}. These fragments assemble into three-dimensional networks,
where edges contain $sp$ motifs and planes can become interlinked through $sp^3$ motifs
(cf. Fig.~\ref{fig:stru_slic}(e-h)), which is the same role played by
interstitial defects
in graphite~\cite{li_2005}.
One of the distinctive characteristics of NP carbon is the typical nanostructuring of
the various tangled graphitic ribbons and graphene sheets, as shown in
Fig.~\ref{fig:stru_slic}. Obviously, the mass density is the main factor determining
the pore sizes, with large open pores observed at low density and tightly packed
graphitic planes interlinked to each other at high density. Due to their importance for
emerging and established applications, pore-size distributions merit further discussion.
We give an in-depth analysis of pore morphology later in this section. 

For all densities, 6-, 5- and 7-membered rings, in that order, are the
main repeat units making up the individual graphene layers as quantified in
Fig.~\ref{fig:ring} through ring statistics. Ring statistics as a measure of
the structural topology is useful to characterize medium-range order and even
to reflect the curvature of the graphitic planes. Figure~\ref{fig:ring} shows
the number of rings normalized per atom, from triangles to dodecagons,
for all four densities studied, in both high-temperature graphitized and
room-temperature NP carbon samples. As a reference point, ideal graphene or
graphite has 0.5 hexagons per atom and zero curvature. Consistent with the
extremely high $sp^2$ fraction reported in Table~\ref{table:sp2}, our results
show a predominance of hexagons over other rings, with an average count of
0.304 (0.299 at 3500 K). The second most common ring motif is pentagons with
0.088 rings/atom (0.086 at 3500 K), followed by heptagons with 0.054 rings/atom
(0.053 at 3500 K). 6-, 5- and 7-membered rings are practically the only ring
motifs contributing to the morphology of NP carbon, with the next most common motif
(octagons) over two orders of magnitude less abundant than hexagons.
Many of the unstable defect rings that remain after 200~ps of graphitization
heal after quenching to 300~K, in particular the 3-membered rings. Ring
distributions are very similar across all four densities except for
larger rings, where differences are only noticeable because of the logarithmic
scale used in Fig.~\ref{fig:ring}. We also note that the small standard deviation
(computed from the different random seeds employed at each density) confirms the
robustness of the simulation protocol. Therefore, we conclude from our results
that ring statistics have no density dependence.

\begin{figure}
    \centering
    \includegraphics[width = \columnwidth]{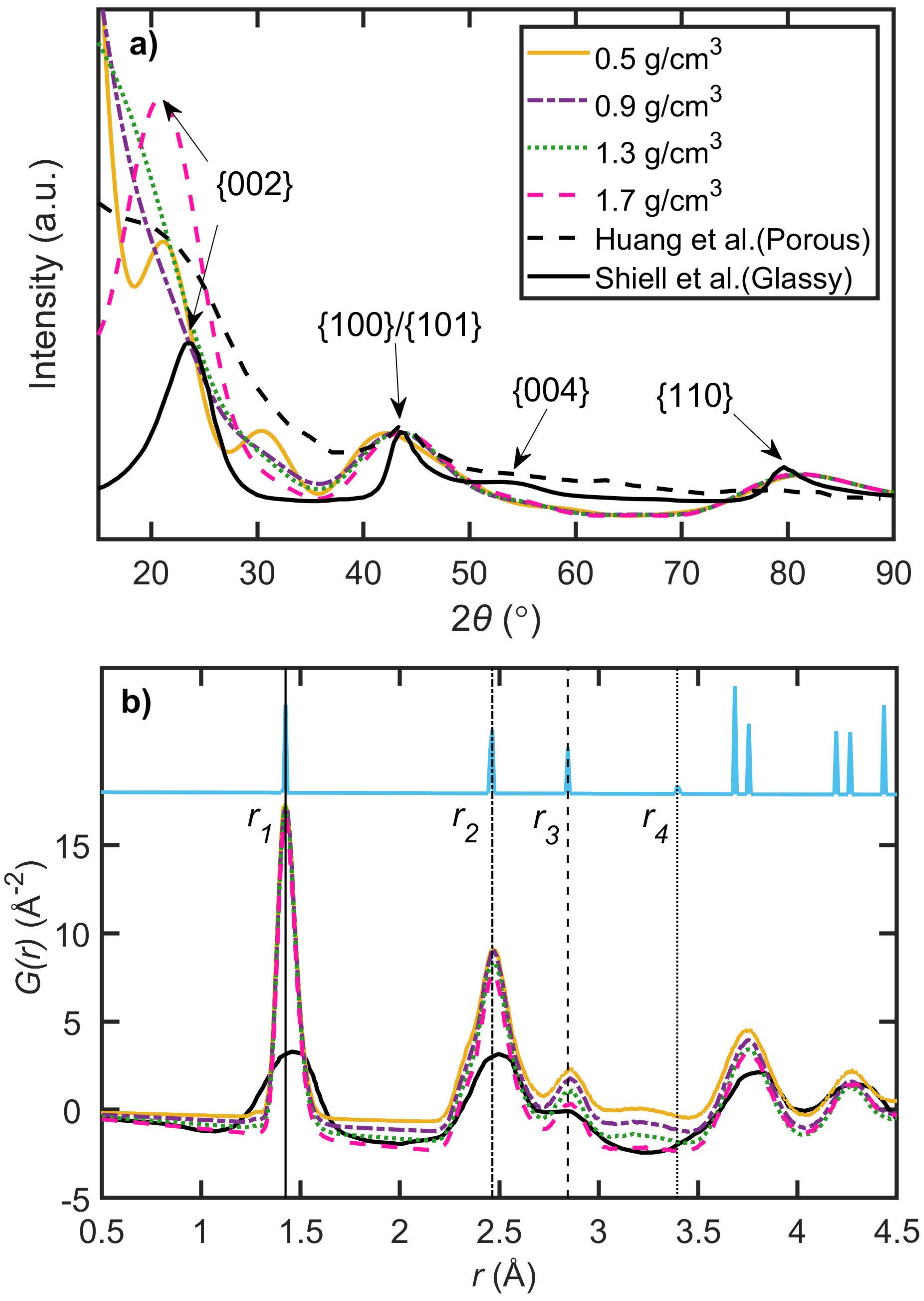}
    \caption{(a) Diffraction patterns and (b) reduced partial distribution
    functions $G(r)$, from our calculations and the experimental results for the
    HTW2500 sample~\cite{Thomas_jncs_2019,neutron_jncs_2021}
    and a NP carbon sample~\cite{huang_2011}. The reflection indices
    are indicated in (a) and the $n$th nearest neighbor distances $r_n$
    ($1\leq n \leq 4$) are represented as vertical lines in (b). For comparison,
    we also present results for graphite computed with $122,880$ atoms and
    lattice parameters of $a=2.46$~\AA{} and $c=6.80$~\AA{}.}
    \label{fig:diffraction_RPDF}
\end{figure}

In order to ensure the validity of the NP carbon structures obtained here, it is of great
importance to assemble the evidence by calculating several key structural and
mechanical indicators and comparing them with experimental data. Among them,
diffraction patterns and radial distribution functions, $G(r)$~\footnote{The reduced
RDF $G(r)$ commonly used in experiment is related to the RDF $g(r)$ most often
reported in simulation work by $G(r) = 4 \pi r \rho_\text{at} (g(r) - 1)$, where
$\rho_\text{at}$ is the atomic density in, e.g., atoms/\AA$^3$.},
are direct tools to characterize the similarity between simulated and experimentally
synthesized structures. For the sake of a robust comparison, the experimental
GC sample HTW2500 reported in Ref.~\citenum{neutron_jncs_2021} and a NP carbon
sample reported in Ref.~\citenum{huang_2011} are chosen.
Figure~\ref{fig:diffraction_RPDF} presents (a) the intensity of X-ray
diffraction (XRD) and (b) the $G(r)$ of the reference experimental
samples and our simulated NP samples. We take the 1.7 g/cm$^3$ model structure
as the closest to HTW2500, because of the similarity in mass density. The labeled
peaks in Fig.~\ref{fig:diffraction_RPDF}(a) are indexed according to the graphite
peaks. The HTW2500 sample has three main peaks at $2\theta \sim 24$, $44$ and
$80$\degree, which correspond to the graphite \{002\}, \{100\} and
\{110\} reflections. For the lower density model structures, no \{002\} reflection
appears due to lower irregular overlapping caused by its more open and diffuse
distribution of graphene sheets.

The interlayer separation can be estimated from Bragg's law according to the
relation for the $l$th-order reflection:
\begin{align}
    d_{00l} = \frac{l \lambda}{2 \sin(\theta_{00l})},
    \label{eq:bragg}
\end{align}
where $d_{00l}$ gives the interplane separation, $\lambda$ is the
wavelength of the X-ray light source and $\theta_{00l}$ is the diffraction
half angle at which the $\{00l\}$ peak is observed.
The shift toward smaller
scattering angles in Fig.~\ref{fig:diffraction_RPDF}(a) as the density decreases
is a clear sign of increasing interlayer spacing, as expected from
Eq.~(\ref{eq:bragg}). Only our computational NP sample of 1.7~g/cm$^3$ shows
a clearly defined peak for the $\{002\}$ reflection. From this peak position,
$\theta \approx 21^\circ$, we can infer an average interlayer spacing of around
4.2~\AA{}. There is no shift in the position of the $\{100\}$ reflection, indicating
an in-plane lattice parameter $a$ of around 2.43~\AA{} (or 1.40~\AA{} interatomic
distance) for $2 \theta_{100} \approx 43^\circ$.

The typical size of crystallites can also be inferred from the XRD
patterns with the Scherrer equation~\cite{scherrer_1912}:
\begin{align}
    L = \frac{K \lambda}{B(2\theta) \cos(\theta)},
\end{align}
where $K \approx 1$ is the Scherrer constant and $B(2\theta)$ is the FWHM
(in radians) of a
given peak on the XRD pattern. The size of in-plane ``crystalline'' pockets
can be inferred from the data for all densities, and it is $L_a \approx 4$~\AA{}
for all of them. This number gives an idea of the typical regions in a graphitic
sheet free of defects, i.e., made of hexagonal rings only. This quantity is in
good agreement with the experimental porous sample which we compare to ours in
Fig.~\ref{fig:diffraction_RPDF}. The size of crystallites along the direction of
stacking can only be inferred for our 1.7~g/cm$^3$ NP sample, as
$L_c \approx 9$~\AA{}. Therefore, in this sample the graphitic sheets are stacked
\textit{locally} similarly to a graphite crystal within length scales encompassing about three
layers.

\subsection{Pore-size distribution}

\begin{figure}
    \centering
    \includegraphics[width = \columnwidth]{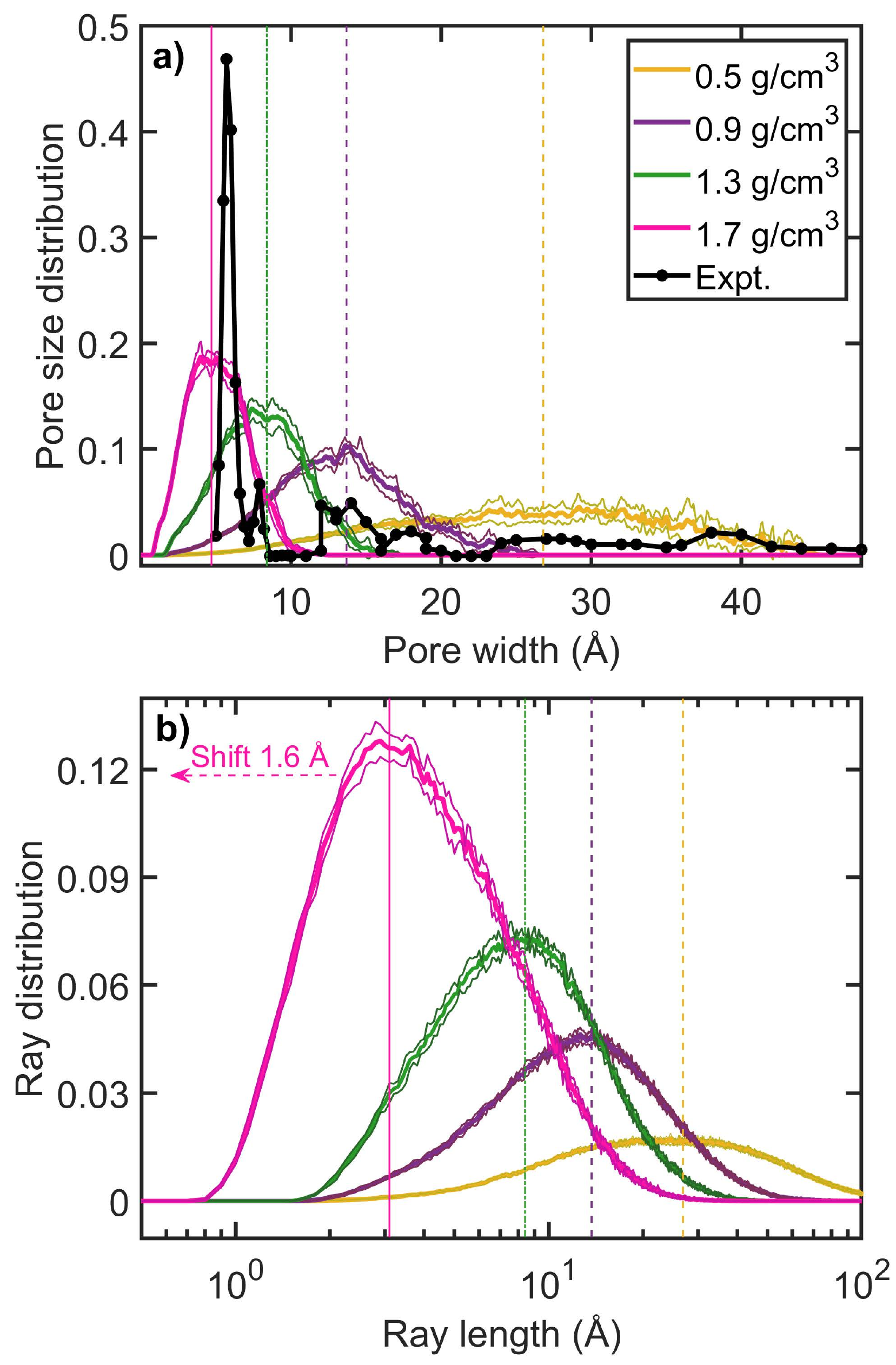}
    \caption{(a) Distribution of pore size as a function of pore width.
    Vertical lines mark diameters of 4.7 in pink, 8.4 in green, 13.7 in purple
    and 26.8~\AA{} in brown for the decreasing order of densities.
    (b) Distribution of the number of rays as a function of ray length. Vertical
    lines mark the ray lengths of 3.1 in pink,  8.4 in green, 13.7 in purple and
    26.8~\AA{} in brown for the decreasing order of densities. Standard deviations
    are provided over six random seeds in the simulations.}
    \label{fig:pore_dist}
\end{figure}

Porosity is one of the most significant properties of NP carbons and
the one that determines their usefulness for many technological applications.
Therefore, the characterization of porosity and pore structures in our
computational samples is a central part of this work. The normalized pore-size
distribution (PSD) and ray-trace histogram are both plotted in
Fig.~\ref{fig:pore_dist}. Comparison to experimental observations is challenging
because of the difficulty in directly measuring PSD in experiments. In
Fig.~\ref{fig:pore_dist}(a), the PSD histograms behave similarly to normal
distributions. We can clearly see, as expected, that the PSDs
shift to smaller diameters with increasing density, and their peaks appear,
in decreasing order of density, at 0.47, 0.84, 1.37 and 2.68~nm for 1.7,
1.3, 0.9 and 0.5~g/cm$^3$, respectively. As shown in Fig.~\ref{fig:stru_slic},
a more open structure means more available space between graphitic planes and,
consequently, the presence of larger pores. At the same time, as the density
increases the PSDs become sharper (i.e., the individual pores have less variation in size).

It should be pointed out that our PSD analysis is based on a spherical
approximation to pores and highly sensitive to small changes in the pore diameter.
It does not reflect subtle changes in surface texture features such as pore
shape~\cite{zeo_jmgm_2013}. To this end, stochastic ray tracing analysis,
as complementary
information for pore structure determination, is also shown in
Fig.~\ref{fig:pore_dist}(b). Following the trend in PSDs, sharper peaks
appear at smaller ray length for denser samples. In general, the low density
of NP carbon explicitly gives a long trailing ray length, which is mainly
caused by open structures, where some rays can travel freely through
continuously connected pores until reaching the wall of a pore. Their
peaks appear, in decreasing order of density, at 0.31, 0.84, 1.37 and 2.68~nm
ray length for 1.7, 1.3, 0.9 and 0.5~g/cm$^3$, respectively. Especially for
lower densities, the fact that the PSD peak position is almost the same as
that of ray trace indicates that a localized enclosed spherical pore curvature
is the dominant morphological feature in the samples modeled. This
observation agrees with the experimental prediction for the existence of spherical
rather than
channel-like or cylindrical-like void regions in GC~\cite{neutron_jncs_2021}.
This result might be related to the excess of 5-membered rings as compared to
their 7-membered counterparts, because the former leads to positive curvature
preferring to form closed spheres. Note that one of the
most common crystallographic defects in graphene, the Stone-Wells
defect,~\cite{stone_1986} involves the direct substitution of four 6-membered
rings with two 5-membered and two 7-membered rings. 5- and 7-member rings
arising in pairs allow to maintain the planarity of the graphitic sheets,
therefore, an excess of either type of ring will lead to curving of the graphitic
planes. An extreme limit of this is the C$_{60}$ fullerene, where only 5- and
6-membered rings are present. Moreover, the peak of ray distribution
for the 1.7 g/cm$^3$ sample shifts slightly (by 1.6~\AA{}) to its corresponding PSD
value, which is due to the local interstitial space between graphitic planes. This
suggests the presence of some pockets of crystallites, i.e., oriented stacked
graphitic sheets, in the 1.7 g/cm$^3$ sample. This is also evidenced by the
\{002\} reflection in the XRD patterns and its size evaluation via the Scherrer
equation above.

Finally, we note that, as in the case for ring counts, the good agreement
(standard deviations visible only at the peak of the PSDs) between the six
individual random-seed runs for each density gives confidence in the robustness
of the simulation protocol.

\subsection{Elastic properties}

Under realistic conditions materials are subjected to different
levels of mechanical stress. If the material's response to stress
is poorly understood, this can lead to unexpected or unwanted effects,
or even to catastrophic consequences such as material failure. Thus,
understanding the mechanical properties of disordered graphitic carbons is
of central importance with regard to their technological and industrial
applications. Therefore, we have evaluated the elastic properties of our
simulated NP carbon samples and characterized the density
dependence of bulk, shear and Young's moduli. Six independent random
seeds for each mass density were used to obtain good statistics for
the elastic moduli of NP carbon. To assess the degree of isotropy in our
samples, we computed the triclinic stiffness tensor of the material, which
has 36 independent components, and then performed an isotropic projection.
This projection is carried out according to the Moakher and Norris (MN)
method~\cite{moakher_2006} with rotation optimization, as implemented in the Mattpy
code~\cite{ref_mattpy,caro_2014b,caro_2015d}. In addition to the ability to
carry out a projection, the MN method provides a quantitative measure of the
similarity between the original tensor and its projection based on
Euclidean distances. In this way, we can establish the degree to which
finite-size effects, on the one hand, and anisotropy, on the other, are present
in our samples. Finite-size effects are a consequence of the limited number
of atoms in the simulation, whereas elastic anisotropy in graphitic carbons
car arise if there is a preferential orientation of the graphitic planes,
as occurs in crystalline graphite.

The relations between the elastic constants that are fulfilled by isotropic
materials are:
\begin{align}
    C_{11} = C_{22} = C_{33};
    \qquad
    C_{12} = C_{13} = C_{23};
    \nonumber \\
    C_{44} = C_{55} = C_{66};
    \qquad
    C_{44} = (C_{11} - C_{12})/2.
    \label{eq:isotropy}
\end{align}
These relations are only valid for three-dimensional isotropy.
When a material displays only in-plane isotropy, only the shear,
axial and biaxial elastic constants \textit{along that particular plane}
satisfy this relation. For instance, in crystalline graphite the
elastic isotropy relations for $c$-plane isotropy are expressed by
$C_{11} = C_{22}$ and $C_{66} = (C_{11} - C_{12})/2$.

The bulk ($B$), shear ($G$) and Young's ($E$) elastic moduli, which are more
commonly reported in the materials and engineering literature than the $C_{ij}$,
are given for isotropic materials as a function of the elastic constants as follows:
\begin{equation}
B = (C_{11} + 2C_{12})/3;
    \label{eq:bulk_modulus}
\end{equation}
\begin{align}
G = C_{44},
 \label{eq:shear_modulus}
\end{align}
and
\begin{equation}
    E = 1/S_{11} = 9BG/(3B+G),
    \label{eq:Young's_modulus}
\end{equation}
where the compliance matrix {\bf S} is obtained by inverting the
$6 \times 6$ matrix of elastic constants {\bf C}.

\begin{figure}[t]
    \centering
    \includegraphics[width = \columnwidth]{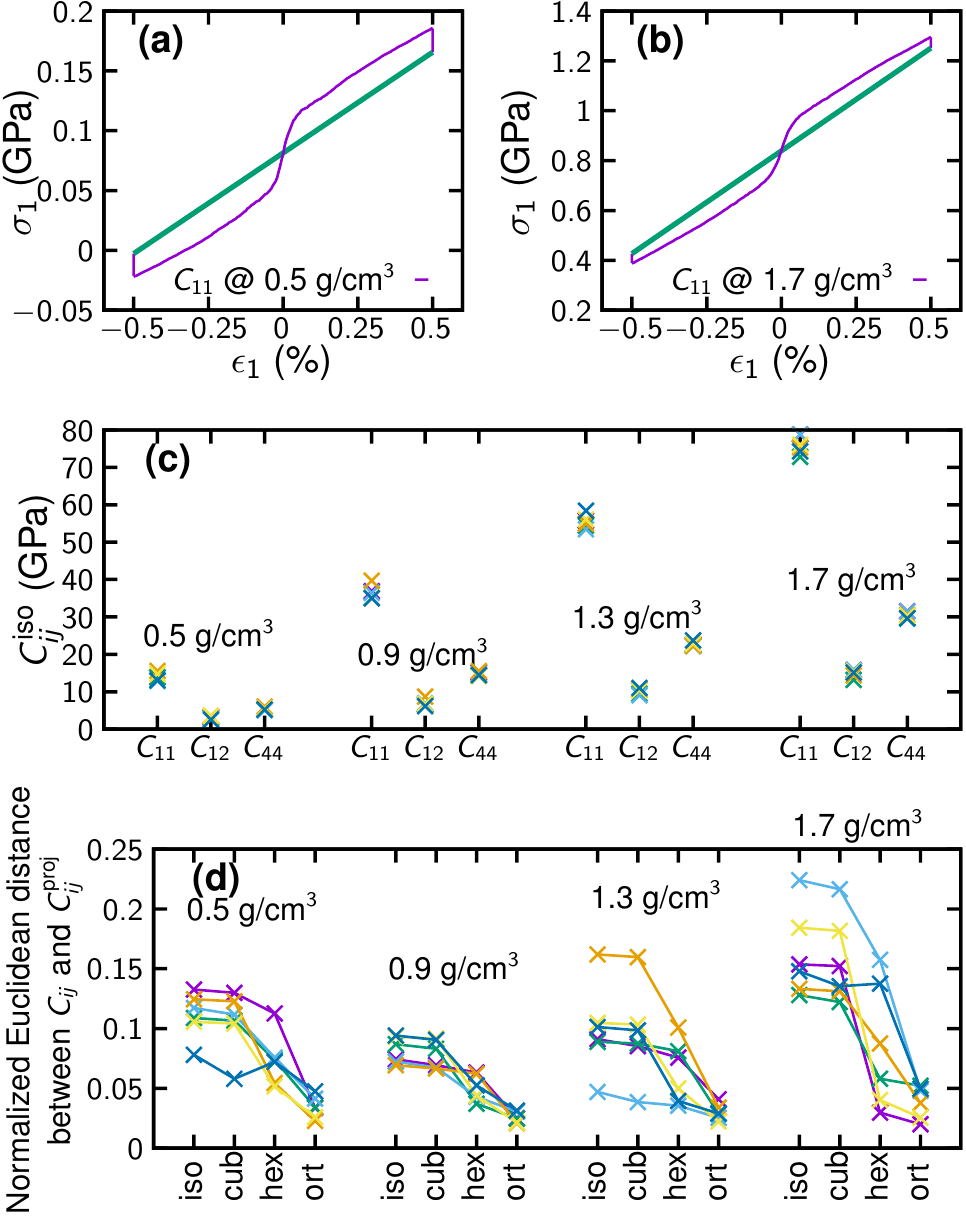}
    \caption{Two examples for 0.5~g/cm$^3$ (a) and 1.7~g/cm$^3$ (b) samples
    for the calculation of $C_{11}$, given by the relation
    between the stress along $x$, $\sigma_1$, and the strain along the same axis,
    $\epsilon_1$. (c) Isotropic projection (which gives $C_{11}$, $C_{12}$ and
    $C_{44}$) of the 36-component $C_{ij}$ tensor computed for all computational
    NP samples. (d) Normalized distance between the triclinic stiffness tensor
    of the samples and the isotropic (iso), cubic (cub), hexagonal (hex) and
    orthorhombic (ort) projections; lines connect the data corresponding to the
    same sample.}
    \label{fig:Cij}
\end{figure}

\begin{figure}[t]
    \centering
    \includegraphics[width = \columnwidth]{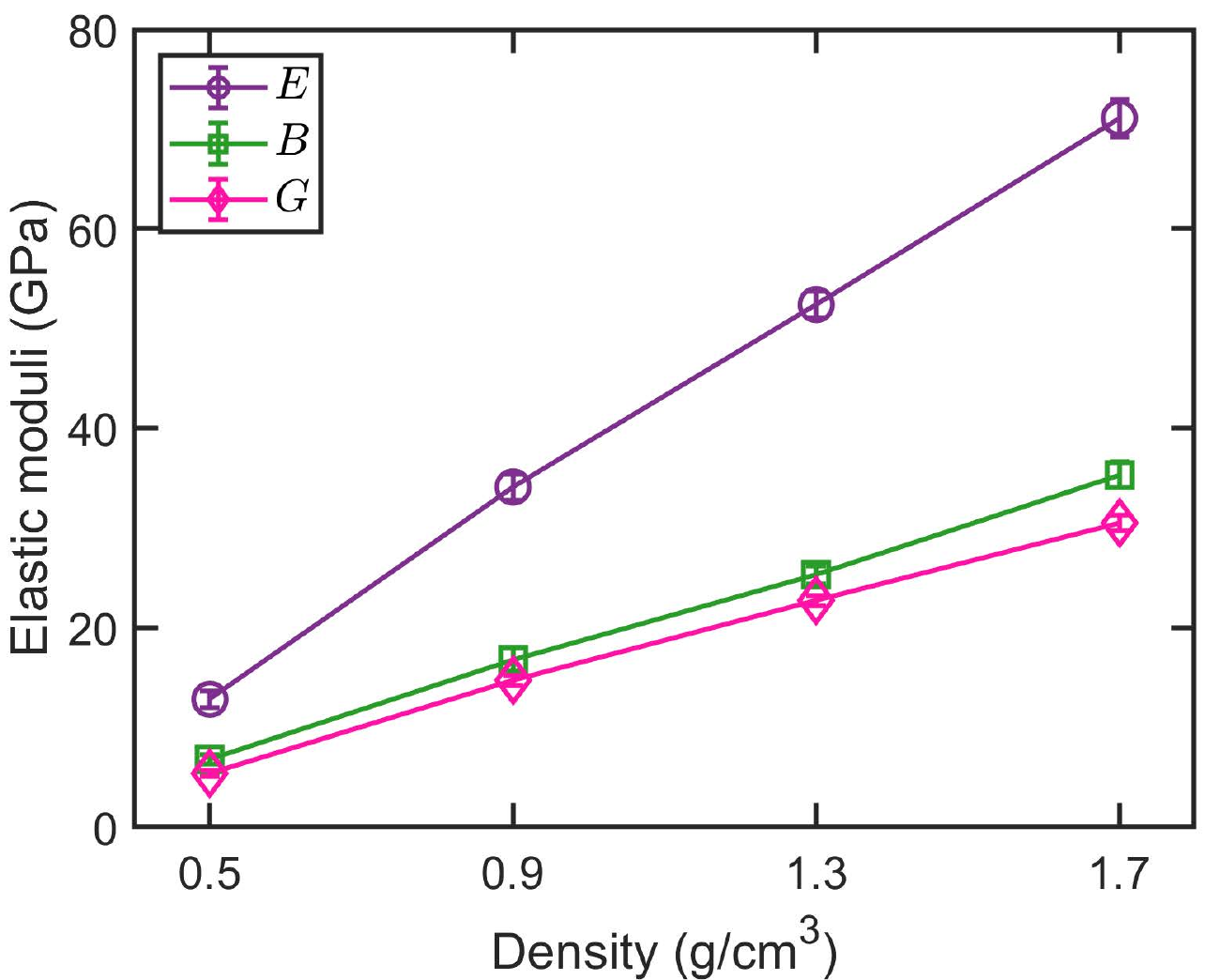}
    \caption{Average Young's ($E$), bulk ($B$) and shear ($G$)
    moduli as a function of density for NP carbons. Standard deviations are given over six random seeds.}
    \label{fig:moduli}
\end{figure}

\begin{table*}[tb]
    \centering
    \begin{tabular}{l c c c c}
    \hline\hline
    & 0.5~g/cm$^3$ & 0.9~g/cm$^3$ & 1.3~g/cm$^3$ & 1.7~g/cm$^3$ \\
    \hline
    $C_{11}^\text{iso}$ (GPa) & $14.0 \pm 0.9$ & $36.4 \pm 1.7$ &
    $55.5 \pm 1.6$ & $76.0 \pm 2.2$ \\
    $C_{12}^\text{iso}$ (GPa) & $3.1 \pm 0.4$ & $6.9 \pm 0.8$ &
    $10.1 \pm 0.7$ & $15.0 \pm 0.9$ \\
    $C_{44}^\text{iso} = G$ (GPa) & $5.4 \pm 0.3$ & $14.7 \pm 0.5$ &
    $22.7 \pm 0.5$ & $30.5 \pm 0.8$ \\
    $B$ (GPa) & $6.8 \pm 0.5$ & $16.8 \pm 1.1$ &
    $25.3 \pm 0.9$ & $35.3 \pm 1.3$ \\
    $E$ (GPa) & $12.8 \pm 0.8$ & $34.1 \pm 1.3$ &
    $52.4 \pm 1.3$ & $71.1 \pm 1.9$ \\
    $C_{11}^\text{hex}/C_{33}^\text{hex}$ & $1.07 \pm 0.21$ &
    $1.05 \pm 0.15$ & $1.00 \pm 0.20$ & $1.41 \pm 0.16$ \\
    \hline\hline
    \end{tabular}
    \caption{Average isotropic elastic constants and elastic moduli, as
    well as ratio of hexagonal axial elastic constants, as a function
    of NP carbon density. The given errors are the standard deviation
    across random seeds in the simulations.}
    \label{tab:cij}
\end{table*}

Because of the complexity of these materials the calculation of the stiffness
tensor is not straightforward. We proceed in the following way. First, the 300~K
structures are further quenched down to close to 0~K to avoid the presence of 
thermal noise when computing the stress tensor. Second, we slowly deform the
simulation box by applying a total of $\pm 0.5$~\% strains over 1~ps in each
direction for all six independent strains,
$\epsilon_j = \epsilon_{1},...,\epsilon_{6}$. This is followed by an energy
minimization with a gradient descent algorithm at the strain end points.
We monitor the six independent
components of the stress tensor $\sigma_i = \sigma_1, ..., \sigma_6$ as a function
of these deformations. This allows us to compute the elastic constants from the
relation $\sigma_i = \sum_j C_{ij} \epsilon_j$ by assuming a linear evolution
as given by the end points. Two examples of these calculations
are given in Fig.~\ref{fig:Cij} (a,b). As can be observed, there is a delayed
elastic response in time, indicated by the non-linear
dependence of the stress on the strain. The vertical lines at the end points
show how the stress is relaxed at fixed strain during the energy minimization step.
The thick straight lines connect the end points of the stress-strain curve, and give
the time-independent elastic response of the material.
In the limit of an infinitely slow deformation the non-linear
curves will approximately converge towards the straight lines.
Third, to estimate the isotropic elastic constants, we carry out an
isotropic projection of the triclinic stiffness tensor by minimizing the
Euclidean distance between the two, constrained to the symmetry of the isotropic
stiffness tensor~\cite{moakher_2006,caro_2014b,ref_mattpy},
which yields the following appropriately weighted averages:
\begin{align}
C_{11}^\text{iso} = & \frac{6}{30} (C_{11}+C_{22}+C_{33}) \nonumber
\\ &+ \frac{2}{30} (C_{12}+C_{13}+C_{23}) \nonumber
\\ &+ \frac{8}{30} (C_{44}+C_{55}+C_{66}),
\\
C_{12}^\text{iso} = & \frac{2}{30} (C_{11}+C_{22}+C_{33}) \nonumber
\\ &+ \frac{4}{30} (C_{12}+C_{13}+C_{23}) \nonumber
\\ &- \frac{4}{30} (C_{44}+C_{55}+C_{66}),
\\
C_{44}^\text{iso} = & \frac{2}{30} (C_{11}+C_{22}+C_{33}) \nonumber 
\\ &- \frac{1}{30} (C_{12}+C_{13}+C_{23}) \nonumber
\\ &+ \frac{6}{30} (C_{44}+C_{55}+C_{66}).
\end{align}
Note that a naive averaging of $C_{11}$, $C_{12}$ and $C_{44}$
does not generally produce a tensor with the correct symmetry.
All the results are shown in Fig.~\ref{fig:Cij} (c). Fourth and last,
we compute the Euclidean distance between the triclinic tensor and tensors
of isotropic, cubic, hexagonal and orthorhombic symmetry (listed in decreasing
order of symmetry), with results shown in Fig.~\ref{fig:Cij} (d). This
approach allows us to establish the best candidate for the underlying symmetry
of each sample.

From Fig.~\ref{fig:Cij} (c), we can observe the smooth evolution of the
elastic constants as a function of mass density, and also appreciate the degree
of variability across samples with the same density, which is small (circa 10~\%)
thanks to the large number of atoms included in these calculations.
Figure~\ref{fig:moduli} shows the average computed for each density, presented
in this case in the form of elastic moduli. All of these values are also
summarized in Table~\ref{tab:cij}.

Going back to Fig.~\ref{fig:Cij} (d) we observe the general expected trend that
as the symmetry of the projection is lowered, the original triclinic tensor can
be better accommodated (the distance between $C_{ij}$ and $C_{ij}^\text{proj}$
decreases). These results also tell us that there is, loosely speaking,
between 5~\% (one of the samples at 1.3~g/cm$^3$) and 23~\%
(one of the samples at 1.7~g/cm$^3$) discrepancy between the triclinic tensor and
its isotropic projection. Some valuable information can be extracted by comparing
the hexagonal projection to both cubic (or even isotropic) and orthorhombic. If
the hexagonal projection is very close to the orthorhombic projection and they
are both relatively far away from the cubic one, there is indication of an
underlying hexagonal elastic response. As we have discussed earlier, the most
characteristic feature in terms of elasticity of hegaxonal lattices is the
existence of an isotropy plane. In other words, these NP samples would show
preferential stacking of graphitic planes along a particular axis. We can
see in Fig.~\ref{fig:Cij} (d) that at low densities the transition from cubic
to hexagonal to orthorhombic projections is smooth, whereas at higher densities
some of the samples show an abrupt transition. This effect can be further
quantified by computing the ratio $C_{11}^\text{hex}/C_{33}^\text{hex}$, where,
as this quantity deviates from unity (usually towards larger values),
the existence of a preferential direction is further emphasized. This effect
is quantified in Table~\ref{tab:cij}, where the highest density samples show
$C_{11}^\text{hex}$ approximately 40~\% larger than $C_{33}^\text{hex}$.
This is indicative of a small
degree of ``graphite-likeness'' in these samples, which is expected to increase
further with density. We note, however, that for our computational NP carbon
samples this ratio is still much smaller (by two orders of magnitude) than that
of crystalline graphite. In addition, the effect is likely due to finite-size
effects, and will probably become even less pronounced for bigger systems. We
therefore conclude that these computational samples are for most practical
purposes elastically isotropic.

\section{Conclusions}

In summary, we have carried out a comprehensive computational study of the atomistic
structure, pore-size distribution and elastic properties of NP carbons
in the mass density range from 0.5~g/cm$^3$ to 1.7~g/cm$^3$. Realistic
simulations of these NP carbon structures, including the modeling of nanopore sizes
up to 40~\AA{} in diameter, have been made possible through the combination
of large simulation cells containing 131,072 atoms each, and the utilization
of an accurate and fast GAP force field for carbon including van der Waals
interactions. The new GAP potential and the generated NP carbon structures
are freely available to the community~\cite{caro_2021,wang_2021b}.

The structural analysis reveals that at all densities studied these NP carbons
are made up mostly of graphitic planes, with circa 97~\% of all atoms being $sp^2$
bonded. Graphitic planes are interlinked via $sp^3$ motifs, whose concentration
increases twofold from 0.5~g/cm$^3$ (0.5~\%) to 1.7~g/cm$^3$ (1.1~\%). Some of the
graphitic planes are terminated with edges involving the presence of $sp$ motifs,
whose concentration follows the opposite trend to that of $sp^3$ motifs (from
3~\% at 0.5~g/cm$^3$ down to 1.4~\% at 1.7~g/cm$^3$). The samples are highly
graphitized, with a topology clearly dominated by 6-membered rings, followed in
smaller quantities by 5- and 7-membered rings, in that order. By comparison, the
presence of smaller or larger rings is negligible, with the next most common ring
motif, 8-membered rings, being more that two orders of magnitude rarer than the 6-membered
rings.

The radial distribution functions for all densities show well-defined peaks
for first, second and third nearest neighbors corresponding to those of
graphite, with the most remarkable feature being a sizable broadening of the
second-neighbors peak. This corresponds to the presence of 5-membered
rings in the samples. The angular distribution function correspondingly shows
a bimodal distribution, with the largest peak centered around 120\degree{}
(hexagonal motifs) and a second smaller peak centered around 108\degree{}
(pentagonal motifs). A small tail also appears for angles larger than
120\degree{} (heptagons and beyond). The calculated XRD patterns show the
expected trends with density and offer a direct route for comparison with
experiment.

The pore-size analysis reveals bell-shaped distributions whose center and
width shift towards larger values as the density decreases. Typical nanopore
diameters increase from 5~\AA{} at 1.7~g/cm$^3$ up to 27~\AA{} at 0.5~g/cm$^3$.
Interestingly, small nanopores are very rare in the low-density NP carbon
samples. This is indicative of homogeneous nanopore size distributions for
a sample of a given density.

Finally, our analysis of elasticity showed that the samples behave
mostly isotropically from a mechanical perspective, with the degree of elastic
isotropy slowly decreasing with density. The elastic moduli of the material
increase monotonically with density, and the Young's modulus increases more
rapidly than bulk or shear moduli. The elastic response of
these NP carbons increased by roughly an order of magnitude from 0.5~g/cm$^3$
to 1.7~g/cm$^3$, with the elastic constants of the latter still more than
an order of magnitude smaller than those of diamond.

We hope that the detailed insight into the atomic structure of NP carbon
materials provided in this work will inspire
and guide future experimental efforts, especially for energy-storage
applications. We also expect that the library of atomic structures provided
to the research community will serve as a starting point for future
computational studies on the electronic, structural and mechanical
properties of NP carbon.

\begin{acknowledgement}
The authors acknowledge funding from the Academy of Finland,
under projects 321713 (M.A.C. \& Y.W.), 330488 (M.A.C.), 312298/QTF
Center of Excellence program ( T.A.-N., Z.F. \& Y.W.), and the China Scholarship Council
under grant no. CSC202006460064 (Y.W.). The authors also acknowledge
computational resources from the Finnish Center for Scientific Computing (CSC)
and Aalto University's Science IT project.
\end{acknowledgement}


\providecommand{\latin}[1]{#1}
\makeatletter
\providecommand{\doi}
  {\begingroup\let\do\@makeother\dospecials
  \catcode`\{=1 \catcode`\}=2 \doi@aux}
\providecommand{\doi@aux}[1]{\endgroup\texttt{#1}}
\makeatother
\providecommand*\mcitethebibliography{\thebibliography}
\csname @ifundefined\endcsname{endmcitethebibliography}
  {\let\endmcitethebibliography\endthebibliography}{}

\end{document}